\documentclass[letterpaper, 10 pt, conference]{ieeeconf}  

\IEEEoverridecommandlockouts                              

\usepackage{graphics} 
\usepackage{verbatim}
\usepackage{epsfig} 
\usepackage{mathptmx} 
\usepackage{url}
\usepackage{times} 
\usepackage{amsmath} 
\usepackage{amssymb}  
\usepackage{physics}
\usepackage{makecell}
\usepackage{xcolor}
\title{\LARGE \bf
 Stability Analysis of the Newton-Raphson Controller for a Class of Differentially Flat Systems
}
\author{K. Niu$^{1}$, Y. Wardi$^{2}$ and C.T. Abdallah$^{3}$
\thanks{$^{1}$School of Electrical and Computer Engineering, Georgia Institute of Technology, Atlanta, GA 30332, USA.
        {\tt\small kniu9@gatech.edu}}%
\thanks{$^{2}$School of Electrical and Computer Engineering, Georgia Institute of Technology, Atlanta, GA 30332, USA.
{\tt\small ywardi@ece.gatech.edu}}%
\thanks{$^{3}$Department of Electrical \& Computer Engineering, Lebanese American University, Beirut, Lebanon, and School of Electrical and Computer Engineering, Georgia Institute of Technology, Atlanta, GA 30318, USA.
        {\tt\small ctabdallah@gatech.edu}}%
}

\begin{document}

\maketitle
\thispagestyle{empty}
\pagestyle{empty}

\begin{abstract}
The Newton-Raphson Controller, established on the output prediction and the Newton-Raphson algorithm, is shown to be effective in a variety of control applications. Although the stability condition of the controller for linear systems has already been established, such condition for nonlinear systems remains unexplored. In this paper, we study the stability of the Newton-Raphson controller for a class of differentially flat nonlinear systems in the context of output regulation and tracking control. For output regulation, we prove that the controlled system is stable within a neighborhood of the origin if the corresponding flat system and output predictor satisfy a verifiable stability criterion. A semi-quantitative analysis is conducted to determine the measure of the domain of attraction. For tracking control, we prove that the controller is capable of driving the outputs to the external reference signals using a specific selection of controller parameters. Simulation results show that the controller achieves regulation and tracking respectively on the inverted pendulum and the kinematic bicycle, suggesting a potential in future control applications.

\end{abstract}

\section{Introduction}
Output regulation and tracking control are two important topics in control systems. To achieve regulation (or tracking), it is desired to design the input to the controlled plant, such that the plant's output could asymptotically approach the origin (or a time-varying reference signal). While control methods for linear systems are well-established, the control for nonlinear systems remains an active area of research. Traditional control approaches including  feedback linearization  and backsteppking \cite{haddad2008nonlinear} guarantee asymptotic stability in the sense of Lyapunov functions, yet both algorithms require specific structures of the controlled plants. Modern control algorithms including Model Predictive Control (MPC) \cite{schwenzer2021review} and Control Lyapunov Function (CLF) \cite{freeman1996control} are shown to be effective in various control applications. However, they often require abundant computational resources and may not have verifiable stability conditions.  

Recently, the authors of this paper proposed the Newton-Raphson Controller, which is shown to be capable of achieving tracking for a number of linear and nonlinear systems \cite{wardi2023tracking}. Two components are involved in the controller, including an output predictor and a Newton-Raphson solver. The predictor estimates the future outputs of the system according to its current states and inputs, and the solver attempts to steer the predictions to the reference signal by adjusting the system inputs using a fluid-flow version of the Newton-Raphson method. Compared to extant algorithms including MPC and CLF, this controller reduces real-time computation loads and may achieve better tracking results \cite{morales2024newton}. Variations of the controller, including model-free control \cite{niu2022model} and consensus control \cite{niu2023consensus}, are also proven to be effective in different control scenarios. Despite the early success, the stability properties of the controller remain unexplored. Although a verifiable stability criterion for linear systems has already been proposed in \cite{wardi2023tracking}, analysis is yet to
be performed for nonlinear systems. To solve this problem, we attempt to find the stability criterion of the Newton-Raphson Controller. While it is difficult to do so for all kinds of systems, we focus on a class of nonlinear dynamics which are \textit{differentially flat}. 

Differential flatness, firstly proposed in \cite{fliess1995flatness}, describes the property of a class of systems that can be transformed to another system (\textit{flat system}) using selected system outputs (\textit{flat outputs}), inputs (\textit{flat inputs}) and their derivatives. Differential flatness has been widely used in tracking \cite{fuchshumer2005flatness}, path planning \cite{zeng2020differential} and safety-critical control \cite{notomista2024safe}. Control design via differential flatness often defines a feedback law in the flat system, which is then transformed back to the original system. Since the flat system is often linear, verification of the system stability is trivial. On the other hand, our approach attempts to define the inputs in the original system and engage the flat system merely as an output predictor, which is different from the flatness based control design method. This definition allows the controller to retain the original Newton-Raphson structure as well as its convergence properties. Moreover, it requires less transformations between the original system and the flat system, avoiding potential computational stress and robustness problems. 

The stability analysis in this paper is divided into two parts, one for output regulation and the other for tracking control. For output regulation, a semi-quantitative analysis reveals the measure of the region of attraction associated with the original controller. For tracking control, we consider a special selection of controller parameters, corresponding to the case where the Newton-Raphson solver runs infinitely fast. Our results show that tracking can be achieved in terms of predictions as long as a verifiable stability condition can be satisfied by the flat system. We simulate the controller using two practical nonlinear systems, including the inverted pendulum and the kinematic bicycle. Simulation results have shown that output regulation and tracking control are achieved on two systems respectively, indicating a potential of this controller for real-world applications.


\section{Tracking Control via Newton-Raphson method}
This section summarizes existing results of the tracking controller based on the Newton-Raphson method \cite{wardi2023tracking}. Consider a nonlinear system defined by:
\begin{equation}
\label{eqn_general_nonlinear}
\begin{split}
    \dot{x}(t) &= f(x(t),u(t))\\
    y(t) &= h(x(t))
\end{split}
\end{equation}
where $x(t) \in \mathbb{R}^{N}$ are the system states, $u(t) \in \mathbb{R}^{M}$ are the system inputs, and $y(t) \in \mathbb{R}^{M}$ are the system outputs. Function $f:\mathbb{R}^N\times\mathbb{R}^M \mapsto \mathbb{R}^{N}$ satisfies the necessary conditions such that the differential equation $\dot{x}(t) = f(x(t), u(t))$ has a unique solution during control, $f(0,0) = 0$, and function $h:\mathbb{R}^{N} \mapsto \mathbb{R}^M$ is continuously differentiable. Output regulation of the nonlinear system (\ref{eqn_general_nonlinear}) requires control inputs $u(t), t \in[0, \infty)$, such that the system outputs satisfy:
\begin{equation}
\label{eqn_regulation_requirement}
    \lim_{t \rightarrow \infty} \sup \Vert y(t)\Vert = 0. 
\end{equation}
Moreover, let $r(t) \in \mathbb{R}^{M}$ be an external reference signal, then tracking control of (\ref{eqn_general_nonlinear}) requires:
\begin{equation}
\label{eqn_tracking_requirement}
    \lim_{t \rightarrow \infty} \sup \Vert y(t) - r(t) \Vert = 0.
\end{equation}
The controller we discuss in this paper attempts to calculate the system input $u(t)$ using variations of the Newton-Raphson method. Generally speaking, the output $y(t)$ of the system (\ref{eqn_general_nonlinear}) is not a function of $u(t)$ but a functional of $x(t_0)$ and $u(\tau), \forall \tau \in [t_0, t)$ for some $t_0 \in \mathbb{R}$. This relationship makes it difficult to calculate $u(t)$ directly according to the requirements (\ref{eqn_regulation_requirement}) or (\ref{eqn_tracking_requirement}) via Newton-Raphson. To overcome this difficulty, the Newton-Raphson controller achieves tracking in terms of predictions. Define the output predictor by:
\begin{equation}
\label{eqn_predictor}
    \tilde{y}(t+T) := g(x(t), u(t), t, T),
\end{equation}
where $T > 0$ is a specified prediction time horizon, and we assume that the prediction function $g(\cdot)$ is continuously differentiable. We will omit $t$ and $T$ in the sequel of this paper if there are no ambiguities, with the predictor being $g(x,u):\mathbb{R}^{N} \times \mathbb{R}^M\mapsto\mathbb{R}^{M}$. Assume that the reference signal $r(t+T)$ is known or can be predicted, then tracking control in terms of prediction indicates:
\begin{equation}
    \label{eqn_prediction_equal}
    \tilde{y}(t+T) := g(x(t), u(t)) = r(t+T).
\end{equation}
Note that the prediction function $g(x(t), u(t), t, T)$ implicitly assumes $\tilde{y}(t+T)$ depends on $x(t)$ and $u(t)$ but not $u(\tau), \tau \in (t, t+T)$. This feature of the prediction function allows a solution (if exists) of Eqn. (\ref{eqn_prediction_equal}) using the continuous Newton-Raphson flow:
\begin{equation}
\label{eqn_nr_dynamics_no_alpha}
    \dot{u}(t) = - \big(\frac{\partial g}{\partial u}(x(t), u(t))\big)^{-1}(g(x(t), u(t)) - r(t+T)),
\end{equation}
assuming that the inversion of $\frac{\partial g}{\partial u}$ exists. As Eqn. (\ref{eqn_nr_dynamics_no_alpha}) converges to the solution asymptotically rather than instantaneously, this control input does not guarantee stability nor tracking. To accelerate the solver as well as improve system stability, consider multiplying the RHS of (\ref{eqn_nr_dynamics_no_alpha}) by a speedup factor $\alpha > 0$. This leads to the definition of the \textit{Dynamical Newton-Raphson Controller} (DNRC):
\begin{equation}
\label{eqn_nr_dynamics}
    \dot{u}(t) = - \alpha\big(\frac{\partial g}{\partial u}(x(t), u(t))\big)^{-1}(g(x(t), u(t)) - r(t+T)).
\end{equation}
It it shown that tracking control of the predicted output can be achieved if the system satisfies the $\alpha$-stability assumption, defined as follows:

\textbf{Definition 1} (\textit{$\alpha$-stability}): Let $\mathcal{C}$ be the set of bounded continuous signals $r(t):\mathbb{R}_+\mapsto \mathbb{R}^M$ with piecewise-continuous and bounded derivatives. The system (\ref{eqn_general_nonlinear}) controlled by (\ref{eqn_nr_dynamics}) is said to be $\alpha$-stable over an open set $ \mathcal{G} \subset L^{\infty}(\mathbb{R}^M)$ and a closed set $\mathcal{F} \subset \mathbb{R}^{N+M}$, if there exists $\bar{\alpha} > 0$, such that for every $\alpha > \bar \alpha$, every reference input $r(t) \in \mathcal{G} \cap \mathcal{C}$ and every initial condition $\zeta_0 := (x(0)^\top, u(0)^\top)^\top \in \mathcal{F}$, the trajectory $\zeta(t) := (x(t)^\top,u(t)^\top)^\top, t \in [0,\infty)$ satisfies:
\begin{enumerate}
    \item The partial derivative $\frac{\partial g}{\partial u}(x(t), u(t))$ is non-singular;
    \item There exist two continuous, monotonously non-decreasing functions $\kappa, \beta:\mathbb{R}^+ \mapsto \mathbb{R}^+$, such that:
    \begin{equation}
        \Vert \zeta\Vert_\infty \leq \kappa(\Vert r \Vert_\infty) + \beta(\Vert \zeta_0\Vert).
    \end{equation}
\end{enumerate}
A Lyapunov-based analysis reveals that under the $\alpha$-stability assumption, there exists a constant number $0\leq\eta<\infty$, such that the tracking error of the system (\ref{eqn_general_nonlinear}) and (\ref{eqn_nr_dynamics}) satisfies:
\begin{equation}
\lim_{t\rightarrow\infty}\sup \Vert \tilde{y}(t+T) - r(t+T) \Vert \leq \frac{\eta}{\alpha}. 
\end{equation}
Since $\eta\not=\infty$, increasing $\alpha$ indicates:
\begin{equation}
    \lim_{\alpha\rightarrow\infty} \lim_{t \rightarrow \infty} \sup\Vert \tilde{y}(t+T) - r(t+T) \Vert = 0.
\end{equation}
In fact, setting $\alpha = \infty$ indicates that the Newton-Raphson solver runs infinitely fast. Consequently, the input to the system will always be the solution of Eqn. (\ref{eqn_prediction_equal}). This leads to another form of the Newton-Raphson Control. Assuming that Eqn. (\ref{eqn_prediction_equal}) has one unique solution, denoted by:
\begin{equation}
    u^{*}(t) = g^{-1}(r(t+T); x(t)),
\end{equation}
then setting $u(t) = u^{*}(t)$ yields:
\begin{equation}
\begin{split}
    \dot{x}(t) &= f(x(t), u^{*}(t)),\\
    y(t) &= h(x(t)).
\end{split}
\end{equation}
We refer to the control input $u(t) = u^{*}(t)$ as the \textit{Statical Newton-Raphson Controller} (SNRC).

We remark that while the SNRC may guarantee stability and smaller tracking error for a broader class of systems, the DNRC, whose controller speed $\alpha$ is finite, may still be favorable in some scenarios. For example, defining $\dot{u}(t)$ rather than $u(t)$ may smooth out discontinuities or oscillations in the reference signal $r(t)$; a dynamical input $\dot{u}(t)$ makes it easier to incorporate Control Barrier Functions (CBFs) \cite{ames2021integral}, and the DNRC may require less computational resources. 

We conclude this section with a brief discussion of the predictor $g(\cdot, \cdot)$. Although the original Newton-Raphson controller does not require specific forms of the predictor $g(\cdot, \cdot)$, our stability analysis relies on a proper selection of the prediction function. One commonly used prediction function is the \textit{fixed-input predictor}. Denote the state of the predictor by $\xi(\tau) \in \mathbb{R}^{N}$ satisfying the following differential equation:
\begin{equation}
    \label{eqn_predictor_dynamics}
    \begin{split}
        \dot{\xi}(\tau) &= f(\xi(\tau), \mu(\tau))\\
        \tilde{y}(\tau) &= h(\xi(\tau)).
    \end{split}
\end{equation}
Setting $\xi(t) = x(t)$, $\mu(\tau) \equiv u(t)$ and simulate the system (\ref{eqn_predictor}) over $\tau \in [t, t+T]$ leads to the solution:
\begin{equation}
    \tilde{y}(t+T) = h(\xi(t+T)),
\end{equation}
which is our desired output prediction.

\section{Stability analysis for Dynamical Newton-Raphson Controller}
In this section we analyze the stability of the Dynamical Newton-Raphson Controller for a class of differentially flat systems. A system $\dot{x} = f(x,u)$ is said to be differentially flat if we can choose $y_f := (y_{f_1}, y_{f_2}, \hdots, y_{fm})$ such that:
    \begin{itemize}
        \item There exist a function $h_f$, s.t. $y_f = h_f(x, u, \dot{u},\hdots, u^{(r)})$
        \item There exist functions $\Theta$, $\Xi$, s.t. $x = \Theta(y_f,\dot{y}_f,\hdots, y^{(p)}_f)$, $u = \Xi(y_f, \dot{y}_f,\hdots, y^{(q)}_f)$.
    \end{itemize}
Specifically, we consider the differentially flat systems with the following properties: 

\textbf{Assumption 1}: The system outputs $y$ of (\ref{eqn_general_nonlinear}) satisfy:
\begin{equation}
    y:= [y_1, y_2,\hdots, y_M]^{\top} = [x_{p_1}, x_{p_2}, \hdots, x_{p_M}]^\top
\end{equation} for some $p_i \in \{1,2,3,\hdots, N\}$.

\textbf{Assumption 2}: The flat output $y_f$ is the same as the original output $y$ for the same system state, namely $y_f(t) = y(t) = h(x(t)), \forall x(t) \in \mathbb{R}^M$.

\textbf{Assumption 3}: There exist flat states $z \in \mathbb{R}^{N}$ and flat inputs $v \in \mathbb{R}^{M}$, such that:
\begin{enumerate}
    \item $y_f(t) = [z_1(t), z_2(t),\hdots, z_M(t)]\ (M \leq N)$ 
    \item The flat system is linear, i.e. 
    \begin{equation}
    \label{eqn_flat_system}
    \begin{split}
        \dot{z} &= Az + Bv\\
        y_f &= h_f(z) = Cz
    \end{split}
    \end{equation} 
    for some $A \in \mathbb{R}^{N\times N}, B \in \mathbb{R}^{N \times M}$.
    \item There exist two invertible and continuously differentiable transformations $\Phi:\mathbb{R}^{N} \mapsto \mathbb{R}^{N}$ and $\Gamma:\mathbb{R}^{N}\times \mathbb{R}^{
 M} \mapsto\mathbb{R}^M
$, such that:
    \begin{equation}
        \begin{split}
            z &= \Phi(x), \Phi(0) = 0, x = \Phi^{-1}(z) := \phi(z);\\
            v &= \Gamma(x, u), \Gamma(0,0) = 0, u = \Gamma^{-1}(x, v) := \gamma(z, v).
        \end{split}
    \end{equation}
\end{enumerate}
While most of the assumptions above are common when applying differential flatness, the transformation $\Phi(\cdot)$ is specific for our system and seems strong. However, it could still be satisfied by a large number of dynamical systems, including inverted pendulums \cite{wikipedia2024pendulum} and kinematic bicycles \cite{kong2015kinematic}. We further assume that the function $\Gamma(\cdot, \cdot)$ and $f(\cdot, \cdot)$ satisfy the following Lipschitz conditions: 

\textbf{Assumption 4}: There exist two compact sets $\mathcal{X} \subset \mathbb{R}^{N}, \mathcal{U}\subset \mathbb{R}^{M}$, such that both sets contain a neighborhood of $0$, and the function $f(x, u)$ and the partial derivative $\frac{\partial \Gamma}{\partial x}(x, u)$ are locally Lipschitz on $\mathcal{X} \times \mathcal{U}$.

We denote the sets in the flat space corresponding to $\mathcal{X}, \mathcal{U}$ by $\mathcal{Z} := \Phi(\mathcal{X}), \mathcal{V} := \Gamma(\mathcal{X}, \mathcal{U})$, whose existences are guaranteed by Assumption 3. 

The stability of the Dynamical Newton-Raphson Controller for output regulation requires a fixed-input predictor derived from the flat system. Fixing the input $v$ in the interval $[t, t+T]$ and solving the linear system $\dot{z} = Az + Bv$ gives:
\begin{equation}
    \tilde{z}(t+T) = e^{AT}z(t) + \int_{0}^{T}e^{A(t-\tau)}d\tau Bv(t) := Rz(t) + Sv(t),
\end{equation}
thus the flat prediction is given by:
\begin{equation}
    g_{f}(z(t), v(t)) := C\tilde{z}(t+T) = CRz(t) + CSv(t).
\end{equation}
With the flat predictor $g_{f}(\cdot, \cdot)$, we have the following results: 

\textbf{Proposition 1}: There exist a positive number $\delta$ and two positive numbers $K_S$ and $K_L$, such that the nonlinear system (\ref{eqn_general_nonlinear}), satisfying Assumption 1-4, and regulated by the Dynamical Newton-Raphson Controller
\begin{equation}
\label{eqn_original_flat_control}
\begin{split}
    \dot{u}(t) &= -\alpha\big(\frac{\partial g_f}{\partial u}(z,v)\big)^{-1}g_f(z,v)\\
    &= -\alpha\big(\frac{\partial g_f}{\partial u}(\Phi(x),\Gamma(x, u)\big)^{-1}g_f(\Phi(x),\Gamma(x,u)),
\end{split}
\end{equation}
is locally asymptotically stable at $y = 0$ over the set $\bar{B}_{\delta}(0) \cap (\mathcal{X}\times\mathcal{U})$, if the flat system $\dot{z} = Az + Bv$ corresponding to $\dot{x} = f(x,u)$ is asymptotically stable at $y_z = 0$ under the control:
\begin{equation}
    \dot{v}(t) = -\alpha\big(\frac{\partial g_f}{\partial v}(z,v)\big)^{-1} g_f(z,v).
\end{equation}
Moreover, the radius of the closed ball $\bar{B}_{\delta}(0)$ satisfies:
\begin{equation}
    \delta < K_S \leq \alpha K_L.
\end{equation}
\textbf{Proof}: First we prove the existence of $\delta $ and $K_S$. Consider the flat system regulated by the Dynamical Newton-Raphson Controller. Using the flat predictor, the system together with the control input becomes:
\begin{equation}
    \begin{split}
        \dot{z} &= Az + Bv\\
        \dot{v} &= -\alpha (CS)^{-1}CRz - \alpha v.
    \end{split}
\end{equation}
Denote $\bar{z}(t) := [z(t)^\top, v(t)^\top]^\top$, the combined flat system is:
\begin{equation}
    \dot{\bar{z}} = \bar{A}\bar{z}, \bar{A} := \begin{bmatrix}
        A&B\\-\alpha (CS)^{-1}CR&-\alpha I.
    \end{bmatrix}
\end{equation}
Since the flat system is stable, there exist two positive definite matrices $P, Q \in \mathbb{R}^{(N+M)\times(N+M)}$, such that:
\begin{equation}
    V(\bar{z}) = \bar{z}^\top P\bar{z} > 0, \dot{V}(\bar{z}) = -\bar{z}^\top Q\bar{z} < 0,  \forall \bar{z} \not = 0.
\end{equation}
Next, we consider the original system (\ref{eqn_general_nonlinear}) controlled by (\ref{eqn_original_flat_control}). With the transformation $\Gamma(\cdot, \cdot)$, the flat input $v$ becomes: 
\begin{equation}
\begin{split}
    \dot{v} &= \frac{d\Gamma}{dt}(x,u) = \frac{\partial \Gamma}{\partial x}\dot{x} + \frac{\partial \Gamma}{\partial u}\dot{u}\\
    &= \frac{\partial \Gamma}{\partial x}\dot{x} -\alpha \frac{\partial \Gamma}{\partial u}(\frac{\partial g_f}{\partial u}(z,v))^{-1}g_f(z, v)\\
    &=\frac{\partial \Gamma}{\partial x}\dot{x} - \alpha\frac{\partial \Gamma}{\partial u}(\frac{\partial g_f}{\partial v}\frac{\partial v}{\partial u})^{-1}g_f(z,v)\\
    &= \frac{\partial \Gamma}{\partial x}\dot{x} - \alpha(\frac{\partial g_f}{\partial v})^{-1}g_f(z,v).
\end{split}
\label{eqn_original_flat_control_in_flat}
\end{equation}
Define
\begin{equation}
    e(z,v) := \frac{\partial \Gamma}{\partial x}(x,u)\dot{x} = \frac{\partial \Gamma}{\partial x}(\phi(z), \gamma(z,v))\dot{\phi}(z).
\end{equation}
and substitute $g_f(\cdot, \cdot)$, the flat system corresponding to the original controlled nonlinear system is:
\begin{equation}
    \dot{\hat{z}} = \bar{A}\hat{z} + \bar{B}e(\hat{z}).
\end{equation}
Here $\hat{z} := (z,v)$ is the combined system states and inputs corresponding to the flat system (\ref{eqn_flat_system}) controlled by (\ref{eqn_original_flat_control_in_flat}), and $\hat{z}$ is different from $\bar{z}$ due to the different definitions of the inputs. $\bar{B} = [\mathbf{0}_{M\times N}, I_{M\times M}]^\top$. Using the Lyapunov function $V$ again, we have:
\begin{equation}
    V(\hat{z}) = \hat{z}^\top P\hat{z}.
\end{equation}
Take derivatives with respect to time:
\begin{equation}
    \begin{split}
        \dot{V}(\hat{z}) &= \dot{\hat{z}}^\top P\hat{z} + \hat{z}P\dot{\hat{z}}\\
        &= (\bar{A}\hat{z} + \bar{B}e(\hat{z}))^\top P \hat{z} + \hat{z}^\top P(\bar{A}\hat{z} + \bar{B}e(\hat{z}))\\
        &= -\hat{z}^\top Q\hat{z} + 2e^{\top}(\hat{z})\bar{B}^\top P\hat{z}\\ &\leq -\hat{z}^\top Q\hat{z} + 2\Vert e^{\top}(\hat{z})\bar{B}^\top P\hat{z} \Vert \\&\leq -\lambda_{\min}(Q)\Vert \hat{z}\Vert^2 + 2\lambda_{\max}(P)\Vert\hat{z}\Vert\Vert\bar{B}e(\hat{z})\Vert.
    \end{split}
\end{equation}
By assumption 4, there exists $L_1, L_2, L_3 > 0$, such that:
\begin{equation}
\begin{split}
\Vert \frac{\partial \Gamma}{\partial x}(x,u)\Vert &\leq L_1\Vert(x,u)\Vert \leq L_1\delta,    \\
\Vert f(x,u)\Vert &\leq L_3\Vert(x,u)\Vert = L_3\Vert(\phi(z),\gamma(z,v))\Vert \leq L_2\Vert\hat{z}\Vert,
\end{split}
\end{equation}
where $\delta$ is the radius of the closed ball in which the trajectory stays. Therefore:
\begin{equation}
\begin{split}
    \dot{V}(\hat{z}) &\leq -\lambda_{\min}(Q)\Vert\hat{z}\Vert^2 + 2 L_1L_2\delta\lambda_{\max}(P)\Vert \hat{z}\Vert^2\Vert\bar{B}\Vert\\&=\Vert\hat{z}\Vert^2(-\lambda_{\min}(Q) + 2L_1L_2\delta\lambda_{\max}(P)\Vert\bar{B}\Vert).
\end{split}
\end{equation}
Let:
\begin{equation}
    K_S = \frac{\lambda_{\min}(Q)}{2L_1L_2\lambda_{\max}(P)\Vert\bar{B}\Vert},
\end{equation}
we have $\dot{V}(\hat{z}) <0 $ for $\hat{z} \not = 0$ if $\delta < K_S$. Hence, the system is locally asymptotically stable, and Eqn. (\ref{eqn_regulation_requirement}) holds for all $(x^\top, u^\top)^\top \in (\mathcal{X}\times \mathcal{U})\cap \bar{B}_{\delta}(0)$.

Now we prove that $K_S$ is upper-bounded by $\alpha$. Note that:
\begin{equation}
\begin{split}
    \lambda_{\min}(Q) = \min_{\Vert \hat{z}\Vert = 1}\hat{z}^\top Q\hat{z} = 2\min_{\Vert\hat{z}\Vert = 1}(-\dot{\hat{z}}^\top P\hat{z}) \\\leq 2(-\dot{\hat{z}}^\top P\hat{z})\big\rvert_{\hat{z} = e_{k}}
\end{split}
\end{equation}
where $e_k$ is the column vector with the $k$-th element being $1$ and other elements being $0$. Choose $k = N+M$, then:
\begin{equation}
    \dot{\hat{z}}\rvert_{\hat{z} = e_{N+M}} = \begin{bmatrix}
        0&0&0&\hdots&0&-\alpha.
    \end{bmatrix}
\end{equation}
Hence:
\begin{equation}
    \lambda_{\min}(Q) \leq 2\alpha P_{0},
\end{equation}
where $P_{0}$ is the last element in the matrix $P$. Since $P$ is positive definite, the following relationship between $P_0$ and the eigenvalue holds:
\begin{equation}
    \dfrac{P_0}{\lambda_{\max}(P)} \leq 1.
\end{equation}
Therefore:
\begin{equation}
    K_S =\frac{\lambda_{\min}(Q)}{2L_1L_2\lambda_{\max}(P)\Vert\bar{B}\Vert} \leq \frac{\alpha}{L_1L_2\Vert\bar{B}\Vert}.
\end{equation}
Let:
\begin{equation}
    K_L = \dfrac{1}{L_1L_2\Vert\bar{B}\Vert},
\end{equation}
then:
\begin{equation}
    K_s \leq K_L\alpha,
\end{equation}
which completes the second part of the inequality. $\blacksquare$

As a special (and very common) case, many flat systems corresponding to the nonlinear systems follow the form of the chains of integrators. For such case, the following stability condition can be verified:

\textbf{Lemma 1}: There exist positive numbers $\alpha_p > 0$, such that the controlled linear system
\begin{equation}
\begin{split}
    \dot{z} &= Az + Bv\\
    \dot{v} &= -\alpha(\frac{\partial g_f}{\partial v})^{-1}g_f(z,v)
\end{split}
\end{equation}
is asymptotically stable for
\begin{equation}
    A = \begin{bmatrix}
        0_{(p-1)\times1} & I_{(p-1)\times(p-1)}\\
        0_{1\times 1} & 0_{1\times(p-1)}
    \end{bmatrix}, B = \begin{bmatrix}
        0_{(p-1)\times1}\\1
    \end{bmatrix},
\end{equation}
$p \in \{1,2,3,4\}$ if $\alpha > \alpha_p$.

We omit the proof for Lemma 1 for the sake of simplicity. Readers are encouraged to check this lemma using the Routh table. Unfortunately, the controller fails for 5th-order integrators, yet 4th-order integrator is often sufficient for most control applications. Combining Lemma 1 and Proposition 1, the following statement holds true:

\textbf{Corollary 1}: The Newton-Raphson controller using the flat predictor is locally asymptotically stable at $y = 0$ if the flat system corresponding to the nonlinear system is a chain of up to 4-th order integrators. 

Also, since all $\alpha$-stable linear systems can be regulated by the Newton-Raphson Controller, we have the following conclusion:

\textbf{Corollary 2}: The Newton-Raphson controller using the flat predictor is locally asymptotically stable at $y = 0$ if the flat system corresponding to the nonlinear system is $\alpha$-stable. \\
A verifiable criterion for $\alpha$-stability can be found in \cite{wardi2023tracking}.

Finally, we discuss the role of the flat system and the regulation error the flat predictor may introduce. Different from traditional flatness-based control scheme, which designs the control inputs $v$ in the flat system, our Newton-Raphson controller defines the input $u$ (or $\dot{u}$ to be exact) in the original system. The functionality of the flat system is restricted to providing the output prediction but not the control input. Meanwhile, the flat prediction is readily obtained once the flat system is derived, avoiding frequent translation between the actual states $x$ and flat states $z$. These features may avoid problems in flatness-based controls including extra computations and lack of robustness. Regarding the error, it is true that the actual output, the prediction using the original system and the prediction using the flat system are different in general. However, this error does not affect the convergence of the system outputs, as we have already seen in Proposition 1.

\section{Stability analysis for Statical Newton-Raphson Controller}
In this section we consider tracking control. Using differential flatness, we obtain a closed-form expression of the system in the flat space. However, the induced nonlinearity $\frac{\partial \Gamma}{\partial x}\dot{x}$  prevents us from the verification of the $\alpha$-stability. Therefore, for tracking control, we consider the Statical Newton-Raphson Controller corresponding to the case where the controller runs infinitely fast ($\alpha = \infty$).  We also assume that the inversion of the prediction function exists:

\textbf{Assumption 5}: There exists one unique solution $u(t) = u^*(t) \in \mathbb{R}^{M}$ that solves the equation
\begin{equation}
\label{eqn_inversion_u}
    g_f(\Phi(x(t)), \Gamma(x(t),u(t))) = r(t+T).
\end{equation}
for all $x(t)\in\mathbb{R}^{N}, u(t)\in\mathbb{R}^M, r(t+T) \in \mathbb{R}^M$.

With the inversion of the prediction as input, the following stability condition holds:

\textbf{Proposition 2}: The nonlinear system (\ref{eqn_general_nonlinear}), satisfying Assumption 1-3, and 5 and controlled by the Statical Newton-Raphson Controller
\begin{equation}
\begin{split}
    u(t) &= g_{f}^{-1}(r(t+T); x(t))\\
    &=\gamma{\big(}\Phi(x(t)), (CS)^{-1}\big(r(t + T) - CR\Phi(x(t))\big){\big)}
\end{split}
\end{equation}
achieves tracking in terms of prediction:
\begin{equation}
    \hat{y}(t+T) = r(t+T),
\end{equation}
for all $r(t)\in\{r(t): \Vert r(t)\Vert_{\infty} < \infty\}$
if the matrix $A - B(CS)^{-1}CR$ is Hurwitz.

\textbf{Proof}: By definition, $u(t)$ is the solution of the equation (\ref{eqn_inversion_u}). Therefore, by Assumption 3, $v(t)$ is the solution of $g_f(z,v) = r(t+T)$:
\begin{equation}
\label{eqn_solution_v}
    v(t) = (CS)^{-1}(r(t+T) - CRz(t))
\end{equation}
Substituting (\ref{eqn_solution_v}) into (\ref{eqn_flat_system}) gives:
\begin{equation}
\label{eqn_flat_solution}
    \dot{z}(t) = (A - B(CS)^{-1}CR)z(t) + B(CS)^{-1}r(t+T).
\end{equation}
For the flat system above, we first check the boundedness of the states and the inputs. Since the matrix $A - B(CS)^{-1}CR$ is Hurwitz, the linear system (\ref{eqn_flat_solution}) is Input-to-State Stable:
\begin{equation}
    \Vert z(t)\Vert \leq \rho(z(0), t) + \sigma(\Vert r(t)\Vert_{\infty}),
\end{equation}
where $\sigma(\cdot)$ is a class-K function and $\rho(\cdot, \cdot)$ is a class-KL function. Due to the boundedness of the input signal, we have $\Vert z(t)\Vert \not = \infty$. By definition, the input and the prediction also satisfies $v(t)\not=\infty, g_f(z(t), v(t))\not=\infty$, respectively, making the Newton-Raphson Controller feasible. Next, since the input $v(t)$ is the solution of $g_f(z(t), v(t)) = r(t+T)$, tracking of the predictions is naturally satisfied. $\blacksquare$

As discussed before, the original prediction $g(x,u)$, the flat prediction $g_f(z,v)$ and the true output $y(t)$ are different in general. However, these errors can be managed by adjusting the prediction time horizon $T$. Define: 
\begin{equation}
    \begin{split}
        e_p(x,u) &:= g(x,u;t, T) - h(x), \\
        e_f(x,u) &:= g_f(\Phi(x),\Gamma(x,u);t,T) - g(x,u; t,T),\\
    \end{split}
\end{equation}
then the following statement holds true:

\textbf{Lemma 2}: Under Assumption 1,2 and 3, there exists four positive numbers $K_P, K_F, T_1, T_2 > 0$, such that the prediction errors of the nonlinear system (\ref{eqn_general_nonlinear}) satisfy: 
\begin{equation}
\begin{split}
\label{eqn_prediction_error_bound}
    \Vert e_p(x,u) \Vert \leq K_PT, \forall T \leq T_1;\\
    \Vert e_f(x,u)\Vert \leq K_FT, \forall T \leq T_2.
\end{split}
\end{equation}
\textbf{Proof}: First, by definition of the output predictors, the predictions equal to the actual output at $T = 0$:
\begin{equation}
    g(x,u;t,0) = g_f(z, v;t,T) = h(x) = h_f(z). 
\end{equation}
Since we assume that the dynamics $f(x,u)$ is locally Lipschitz, there exists a positive number $L_P$, such that:
\begin{equation}
    \Vert g(x, u;t,T) - g(x,u;t,0) \Vert \leq L_PT
\end{equation}
for some $T \leq T_P$. Also, since the flat system is linear, there exists another positive number $L_F$, such that:
\begin{equation}
    \Vert g_f(z,v;t,T) - g_f(z,v;t,0)\Vert \leq L_FT,
\end{equation}
for some $T \leq T_F$. Therefore:
\begin{equation}
\begin{split}
    \Vert e_p(x,u)\Vert &= \Vert g(x,u;t,T) - h(x)\Vert \\
    &= \Vert g(x,u,t,T) - g(x,u,t,0)\Vert \leq L_PT.
\end{split}
\end{equation}
\begin{equation}
    \begin{split}
        \Vert e_f(x,u) \Vert &= \Vert g(x,u;t,T) - g_f(z,v;t,T)\Vert \\
        &= \Vert g(x,u;t,T) - g(x,u;t,0) \\&~~~+ g_f(z,v;t,0) - g_f(z,v;t,T)\Vert\\&\leq \Vert g(x,u;t,T) - g(x,u;t,0)\Vert \\&~~~+ \Vert g_f(z,v;t,0) - g_f(z,v;t,T)\Vert \\&\leq (L_P + L_F)T, \forall T \leq \min(T_P, T_F).
    \end{split}
\end{equation}
Hence, by choosing $K_P = L_P$, $K_F = L_P+L_F$, $T_1 = T_P$ and $T_2 = \min(T_P, T_F)$, the equation (\ref{eqn_prediction_error_bound}) holds. $\blacksquare$

Similar to the case of output regulation, it is relatively easier to verify the stability for cascades of integrators:

\textbf{Corollary 3}: The Newton-Raphson Controller with $\alpha = \infty$ and the flat predictor achieves tracking in terms of prediction if the corresponding flat system is a chain of up to 4-th order integrators.  

We mention that while the Statical Newton Raphson Controller does not require the continuity of the reference signal, jumps and discontinuities in $r(t)$ may lead to discontinuous control input $v(t)$ in practical use. The Dynamical Controller with the dynamically defined input $\dot{u}(t)$, on the other hand, could provide a continuous approximation of the solution $u^*(t)$, which may make it favorable in some specific applications.

\section{Simulation Results}
\begin{figure}
    \centering
    \includegraphics[width=0.3\linewidth]{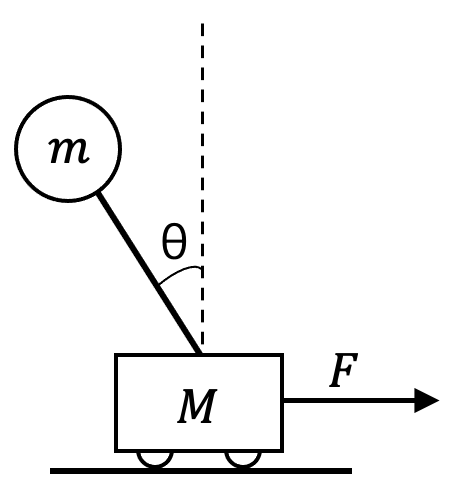}
    \caption{The inverted pendulum, cart-pole model}
    \label{fig_pendulum_structure}
\end{figure}
\subsection{Inverted Pendulum}
First we show the effectiveness of the Dynamical Newton-Raphson Controller for output regulation. The controller is applied on the inverted pendulum (cart-pole) model, which consists of three components: a cart with mass $M$, a weight with mass $m$, and a weightless pole with length $l$. Fig. \ref{fig_pendulum_structure} shows an example of such structure. The objective of the controller is to keep the pole perpendicular to the ground by adjusting the dragging force $F$ applied on the cart. Let $\theta$ be the angle between the pole and the horizontal line, then the dynamics of the inverted pendulum can be described by:
\begin{equation}
\begin{split}
(Ml+ml\sin^2\theta)\ddot{\theta}
+ml\dot{\theta}^2\sin(\theta)\cos(\theta)& \\
-(M+m)g\sin\theta&=F\cos\theta.
\end{split}
\end{equation}
Let $u = F$ be the system input and $y = \theta$ be the system output. Let:
\begin{equation}
    z_1 := \theta, z_2 := \dot{\theta},
\end{equation}
and the flat system becomes:
\begin{equation}
    \begin{split}
        \dot{z}_1 &= z_2\\
        \dot{z}_2 &= \frac{F\cos\theta + (m+M)g\sin\theta -ml\dot{\theta}^2\sin(\theta)\cos(\theta)}{Ml + ml\sin^2\theta}:=v.
    \end{split}
\end{equation}
It can be verified that the original system and the flat system satisfy the aforementioned assumptions for system stability. Let $T$ be the prediction time horizon, and the flat predictor is given by:
\begin{equation}
    g_f(z,v) = z_1 + z_2T + \dfrac{1}{2}vT^2.
\end{equation}
Substituting into Eqn. (\ref{eqn_original_flat_control}) yields:
\begin{equation}
\begin{split}
    \dot{u}(t) = -\alpha\dfrac{2\theta(ml\sin^2\theta + Ml)}{T^2\cos\theta} - \alpha\dfrac{2\dot{\theta}(ml\sin^2\theta + Ml)}{T\cos\theta}\\-\alpha\frac{u\cos\theta + (m+M)g\sin\theta-ml\dot{\theta}\sin(\theta)\cos(\theta)}{\cos\theta}.
\end{split} 
\end{equation}
We test the controller using $T = 0.3s, 0.6s$ and $1.0s$ to show the impact of prediction time horizons. We also start our simulation using different initial conditions, including $\theta(0) = \dfrac{\pi}{6}$ and $\theta(0) = -\dfrac{\pi}{4}$. The controller speedup $\alpha$ is set to $100$ to guarantee system stability as well as improve the rate of convergence. The parameters of the model are selected to be $m = 0.2kg$, $M = 1kg$, $l = 2m$, $g = 9.81m/s^2$. 

\begin{figure}[h]
    \centering
    \includegraphics[width=1.0\linewidth]{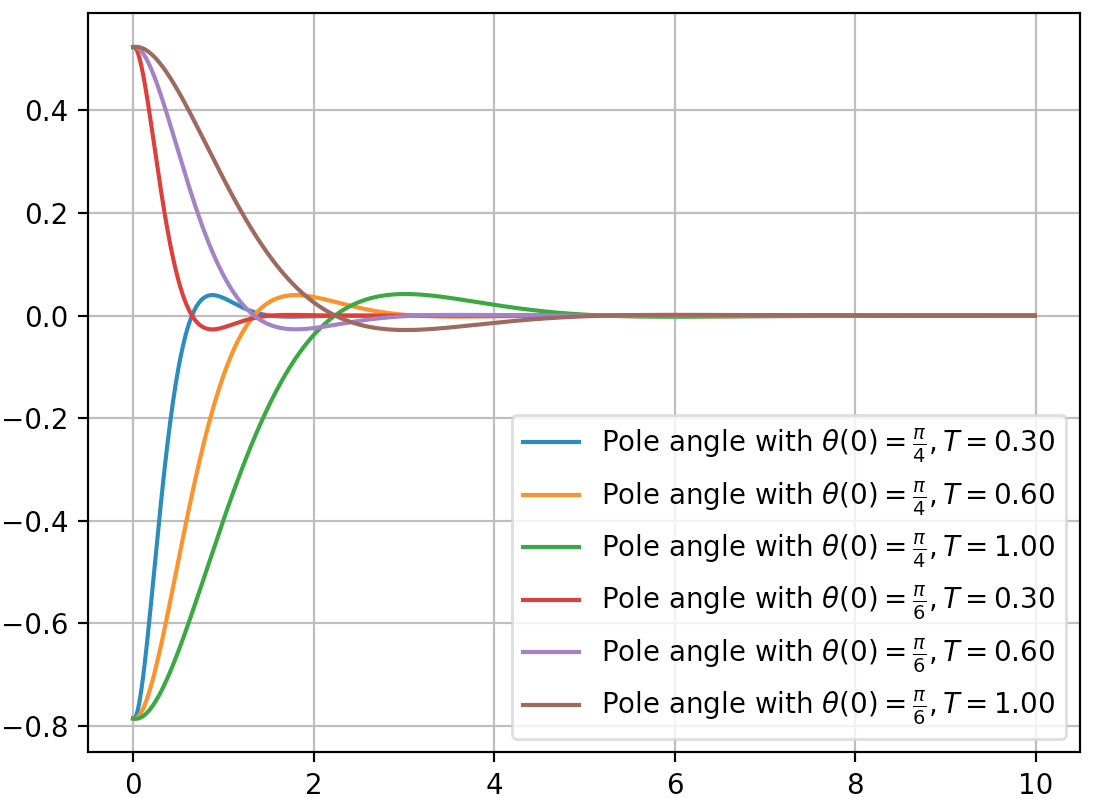}
    \caption{Pole angles of the inverted pendulum}
    \label{fig_pendulum_angle}
\end{figure}
\begin{figure}[h]
    \centering
    \includegraphics[width=1.0\linewidth]{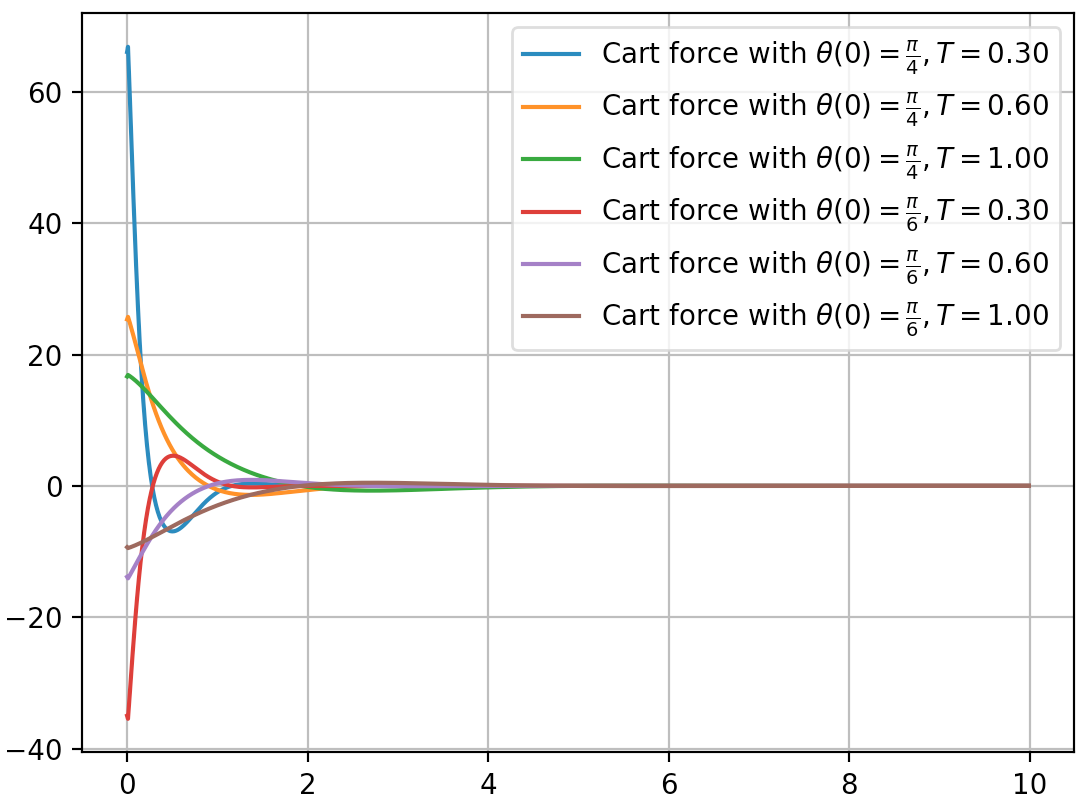}
    \caption{Inputs to the inverted pendulum during simulation}
    \label{fig_pendulum_input}
\end{figure}

The pole angles $\theta$ of the inverted pendulum during simulation are shown in Fig. \ref{fig_pendulum_angle}. Observe that the output converges to $\theta = 0$ for all different cases. To better understand the behavior of the controller, we also plot the system inputs $u$ in Fig. \ref{fig_pendulum_input}. It can be shown that while a smaller time horizon contributes to a faster convergence rate, the input to the system would also become larger. In practical applications, proper parameters should be selected to balance the trade-off between the input and the convergence. 
\subsection{Kinematic Bicycle}
\begin{figure}[b]
    \centering
\includegraphics[width=0.5\linewidth]{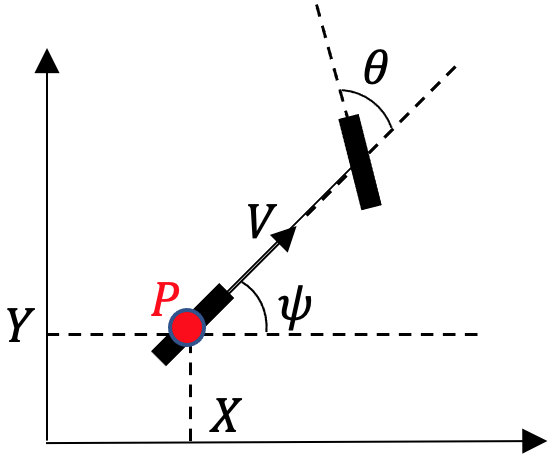}
    \caption{The structure of the kinematic bicycle model}
    \label{fig_bicycle_structure}
\end{figure}
Now, we verify the tracking control of the Statical Newton-Raphson Controller using the kinematic bicycle model. The kinematic bicycle model is a 4-th order nonlinear system representing the dynamics of a vehicle. Assume the length of the vehicle is $L > 0$. Let $P$ be the point located in the middle of the rear axle, and denote the coordinate of $P$ in the 2-D plane by $(X,Y) \in \mathbb{R}^2$. Let $V \in \mathbb{R}$ be the speed of the point $P$, and let $\psi \in \mathbb{R}$ be the vehicle yaw angle, then the state-space model of the vehicle is:
\begin{equation}
\label{eqn_bicycle_dynamics}
    \begin{split}
        \dot{X} &= V\cos(\psi)\\
        \dot{Y} &= V\sin(\psi),\\
        \dot{V} &= a,\\
        \dot{\psi} &= \frac{V}{L}\tan(\theta).
    \end{split}
\end{equation}
The inputs of the system are defined by $u := (a, \theta)^\top$, where $a \in [a_{\min}, a_{\max}]$ is the vehicle acceleration, and $\theta \in [\theta_{\min}, \theta_{\max}]$ denotes the angle between the steering wheel and the vehicle frame. The outputs of the system are defined by $y:= (X, Y)^\top$, and the states of the system are denoted by $x := (X, Y, V, \psi)$. The relationship between the inputs and system states is depicted in Fig. \ref{fig_bicycle_structure}.

To establish the flat system, let $z_1 := y_1 = X$, $z_2 := y_2 = Y$. Taking derivatives of the flat states yields:
\begin{equation}
\label{eqn_bicycle_flat_transform}
\begin{split}
    \dot{z}_1 &= \dot{X} = V\cos(\psi) := z_3 \\
    \dot{z}_2 &= \dot{Y} = V\sin(\psi) := z_4\\
    \dot{z}_3 &= \dot{V}\cos(\psi) -V\sin(\psi)\dot{\psi}\\&=a\cos(\psi) - \frac{V^2}{L}\sin(\psi)\tan(\theta) := v_1\\
    \dot{z}_4 &= \dot{V}\sin(\psi) + V\cos(\psi)\dot{\psi}\\&=a\sin(\psi) + \frac{V^2}{L}\cos(\psi)\tan(\theta) := v_2.
\end{split}
\end{equation}
\begin{figure}[h]
    \centering
    \includegraphics[width=1.0\linewidth]{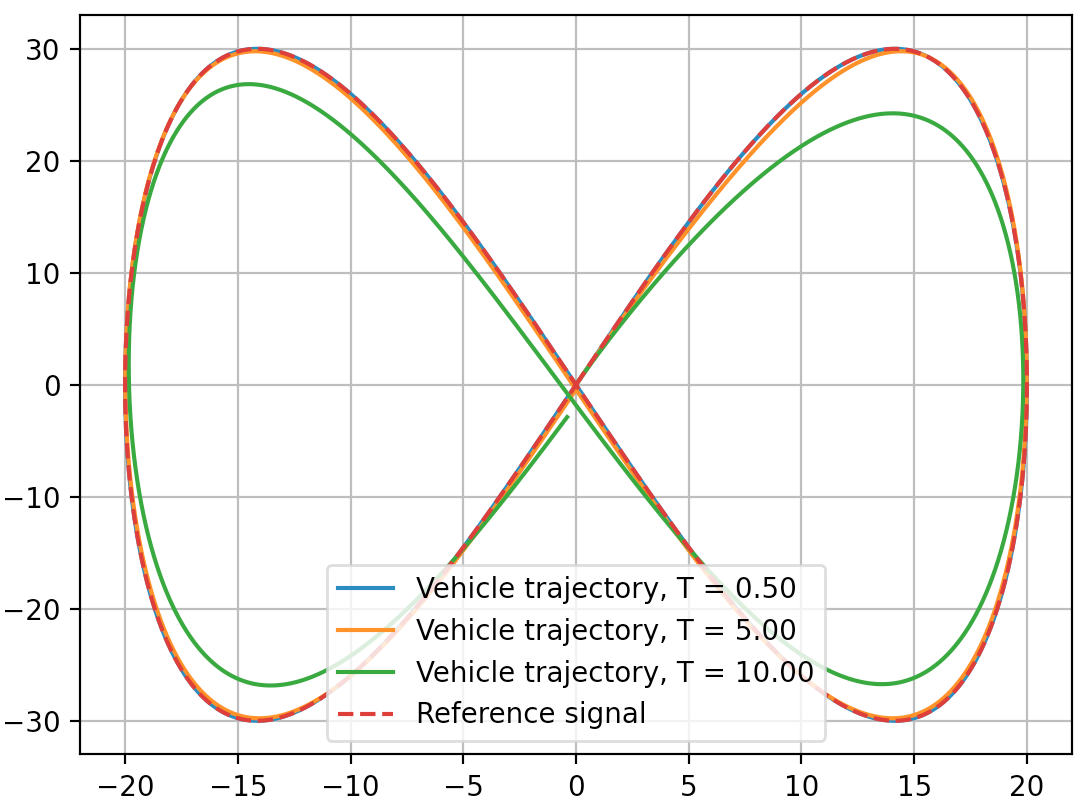}
    \caption{The trajectory of the kinematic bicycle}
    \label{fig_vehicle_trajectory}
\end{figure}

It can be verified that the transformation is invertible if and only if $V \not = 0$. Hence, we assume that the vehicle does not stop during the control period. The flat prediction using the flat system is:
\begin{equation}
    g(z, v) := \begin{bmatrix}
        \hat{z}_1\\ \hat{z}_2
    \end{bmatrix} =  \begin{bmatrix}
        z_1 + z_3T + \frac{1}{2}v_1T^2\\
        z_2 + z_4T + \frac{1}{2}v_2T^2
    \end{bmatrix}.
\end{equation}
\begin{figure}[h]
    \centering
    \includegraphics[width=1.0\linewidth]{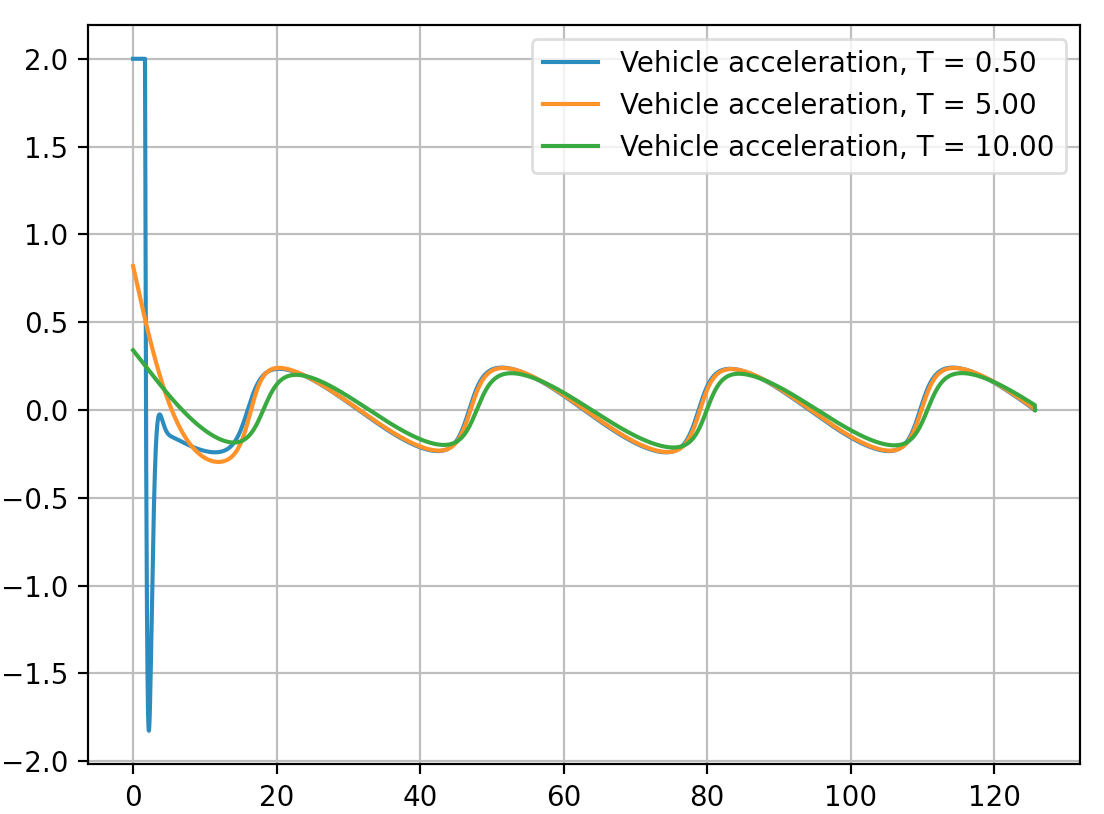}
    \caption{The acceleration of the vehicle}
    \label{fig_vehicle_acceleration}
\end{figure}
Letting $g(z,v) = (r_1, r_2)^\top$ yields the flat inputs:
\begin{equation}
\begin{split}
    v_1 = \frac{2}{T^2}(r_1 - z_1 - z_3T)\\
    v_2 = \frac{2}{T^2}(r_2 - z_2 - z_4T),
\end{split}
\end{equation}
and the actual input $u$ can be readily obtained using Eqn. (\ref{eqn_bicycle_flat_transform}). We simulate the system using the following input constraints: $a \in [-5, 2]\ m/s^2$, $\theta \in [-\frac{\pi}{4}, \frac{\pi}{4}]\ rad$. the external reference signal is:
\begin{equation}
    r(t) = \begin{bmatrix}
        20\sin(0.05t)\\30\sin(0.1t)
    \end{bmatrix}.
\end{equation}

\begin{figure}
    \centering
    \includegraphics[width=1.0\linewidth]{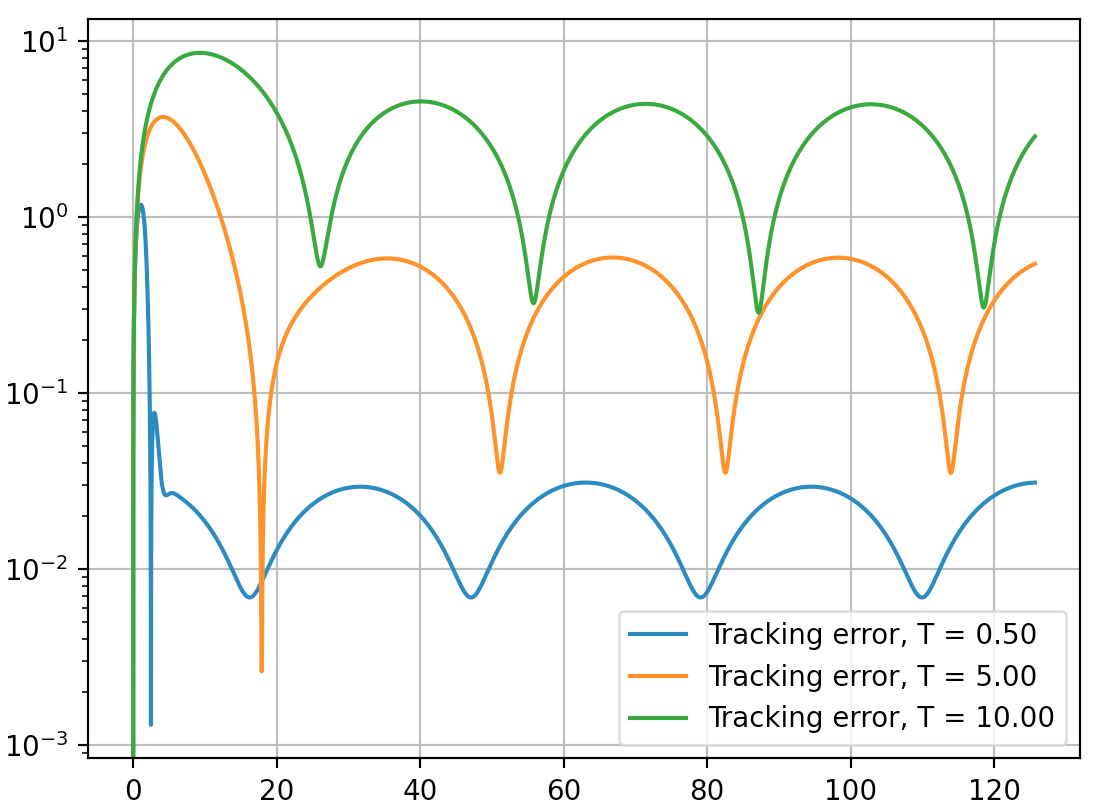}
    \caption{Tracking error of the kinematic bicycle model}
    \label{fig_vehicle_error}
\end{figure}

To illustrate the impacts of the prediction time horizon $T$, we perform three experiments using $T = 0.5s, 5.0s$ and $10.0s$. The trajectories and the accelerations of the vehicle during simulation are shown in Fig. \ref{fig_vehicle_trajectory} and Fig. \ref{fig_vehicle_acceleration} respectively. On the one hand, as the prediction time horizon gets larger, the tracking error also gets larger. On the other hand, larger time horizon also reduces the input transients and avoids input saturation. The tracking error, defined by $e(t) := \Vert y(t) - r(t)\Vert$, is shown in Fig. \ref{fig_vehicle_error}. With $T = 0.5$, the vehicle achieves a tracking error of less than $3cm$, making this controller feasible in real-world autonomous driving scenarios.

\section{Conclusions}
In this paper we investigated the stability properties of the Newton-Raphson controller. The controller is constructed on an output predictor and a Newton-Raphson solver, and the flat system is applied by the predictor to provide an estimated future outputs. We have shown that for output regulation problems, the controller is locally stable if the flat system satisfies a stability criterion. The measure of the region of attraction is estimated using a semi-quantitative analysis. For tracking control, we considered the special case where the Newton-Raphson solver runs infinitely fast. It can be shown that tracking control in terms of output prediction is achieved if the flat system satisfies another stability condition. We also performed simulations to show the effectiveness of the controller using two practical systems.

\end{document}